\documentclass[a4paper,11pt]{article}
\usepackage{jheppub} 

\usepackage{jheppub}
\usepackage{amsmath}
\usepackage{amssymb}
\usepackage{appendix}
\usepackage{multirow}
\usepackage{rotating}
\usepackage{diagbox}
\usepackage{slashed}
\usepackage{float}
\usepackage[dvipsnames]{xcolor}
\usepackage{cancel}
\usepackage{tabularx}
\usepackage{bbold}
\usepackage[capitalize]{cleveref}
\usepackage[normalem]{ulem}
\usepackage{soul}

\makeatletter
\g@addto@macro\bfseries{\boldmath}
\makeatother

\newcommand\supsetsim{\mathrel{\substack{\textstyle\supset\\[-0.2ex]\textstyle\sim}}}

\title{Deconstructing flavor anomalously}

\author[a]{Javier Fuentes-Mart\'in,}
\affiliation[a]{Departamento de Física Teórica y del Cosmos, Universidad de Granada, 18071 Granada, Spain}
\author[b]{Javier M. Lizana}
\affiliation[b]{Physik-Institut, Universit\"at Z\"urich,
Winterthurerstrasse 190, 8057 Zurich, Switzerland}
\affiliation[b]{Instituto de F\'isica Te\'orica UAM/CSIC, Nicolas Cabrera 13-15, 28049 Madrid, Spain}
\emailAdd{javier.fuentes@ugr.es}
\emailAdd{jmlizana@ift.csic.es}

\abstract{
Flavor deconstruction refers to ultraviolet completions of the Standard Model where the gauge group is split into multiple factors under which fermions transform non-universally. We propose a mechanism for charging same-family fermions into different factors of a deconstructed gauge theory in a way that gauge anomalies are avoided. The mechanism relies in the inclusion of a strongly-coupled sector, responsible of both anomaly cancellation and the breaking of the non-universal gauge symmetry. As an application, we propose different flavor deconstructions of the Standard Model that, instead of complete families, uniquely identify specific third-family fermions. All these deconstructions allow for a new physics scale that can be as low as few TeV and provide an excellent starting point for the explanation of the Standard Model flavor hierarchies.
}

\preprint{IFT-UAM/CSIC-24-21}
\makeatletter
\gdef\@fpheader{}
\makeatother

\begin{document}
\maketitle
\flushbottom

\section{Introduction}
\label{sec:intro}

Flavor universality of the Standard Model (SM) gauge sector could be an emergent property below the TeV scale. The possibility of having new physics (NP) that extends the SM gauge sector in a flavor non-universal manner opens up an interesting path to explain the SM flavor hierarchies while providing a rich phenomenology. Although flavor observables like meson mixing put strong bounds on non-universal physics between first and second families, pushing the scale to the PeV range, the scale of universality breaking between third and light families can be as low as few TeV~\cite{Allwicher:2023shc}. 

This hierarchy of energies suggests that flavor can be addressed by having multiple NP scales~\cite{Berezhiani:1992pj,Barbieri:1994cx,Dvali:2000ha,Panico:2016ull}, something that is achieved by {\it deconstructing flavor}. In analogy to the deconstruction of an extra dimension~\cite{Georgi:1985hf,Arkani-Hamed:2001kyx,Hill:2000mu}, we can extend some of the factors of the SM gauge group into identical products that break to their diagonal subgroup at low-energies. Different SM fermions can then be charged under the different factors, hence resulting in violations of flavor universality. Whereas it is phenomenologically viable to break universality between first and second families above the PeV scale, these models are effectively described around the TeV scale by a gauge group that charges the light families universally and the third family differently. Examples of models deconstructing some of the SM factors, or its UV completions, can be found in~\cite{Chivukula:2013kw,Bordone:2017bld,Greljo:2018tuh,Davighi:2023iks,FernandezNavarro:2023rhv,Davighi:2023evx,FernandezNavarro:2023hrf,Davighi:2023xqn,Barbieri:2023qpf,Capdevila:2024gki}. Among them, a particular interesting realization is the so-called 4321 models~\cite{Bordone:2017bld,Greljo:2018tuh}, a color deconstruction based on the $\mathcal{G}_{4321}\equiv SU(4)\times SU(3)^\prime\times SU(2)_L\times U(1)_X$ gauge group. This class of models has been extensively studied~\cite{Cornella:2019hct,Cornella:2021sby,Crosas:2022quq,Allwicher:2023aql,Haisch:2022afh}, as it provides one of the most compelling explanations for the observed hints of deviations from the SM predictions in $B$ decays~\cite{Capdevila:2023yhq,Koppenburg:2023ndc}. While these hints have become less significant nowadays, 4321 models present multiple theoretically-appealing features that go beyond the explanation of these anomalies. For instance, they allow for low-scale unification of third-family quarks and leptons \`a la Pati-Salam~\cite{Pati:1974yy} and provide a natural explanation for the smallness of the Cabibbo-Kobayashi-Maskawa (CKM) mixing with the third family. 

In general, flavor-deconstructed models require a sector responsible for the breaking of the extended symmetry to the SM gauge group. A common approach consists in the inclusion of {\it link} scalar fields: fields charged under two group factors that spontaneously break them to their diagonal once they acquire a vacuum expectation value (vev). In such scenarios, if there are no other fermions than those in the SM, the cancellation of gauge anomalies requires to charge full SM generations under each group factor.\footnote{A possibility to break this condition is to include {\it anomalons}, chiral fermions of the UV symmetry that get mass from the breaking and become vector-like fermions under the SM gauge group. Examples of similar constructions can be found in~\cite{Davighi:2021oel,Antusch:2023shi,Capdevila:2024gki}.} 
This imposes some rigidity on the construction of these models that, in certain situations, could be interesting to relax. A radiatively stable alternative for the symmetry breaking, analogous to technicolor theories for electroweak (EW) symmetry breaking, is to add an {\it hyper-sector} with strong dynamics that develops condensates breaking the UV symmetry. This avoids the inclusion of extra ad-hoc scalars. Furthermore, if the composite sector responsible of the lowest breaking confines at the TeV scale, it could be connected to composite Higgs solutions of the hierarchy problem, similarly to the 4321 model presented in~\cite{Fuentes-Martin:2020bnh}. 

In this article, we explore the possibility of charging new fermionic degrees of freedom of an hypothetical strongly-coupled sector so we can split the SM fermions among different factors of a deconstructed SM symmetry while avoiding gauge anomalies.
Although these fermionic fields are chiral and have to be massless, the strong dynamics ensures that they do not appear as asymptotic states but only as partons of new hyper-baryons of the composite sector at the confinement scale. A variant of this idea is actually realized in nature: the breaking of $SU(N_f)_L\times SU(N_f)_R\times U(1)_{B-L}\to SU(N_f)_V\times U(1)_{B-L}$ triggered by the quark condensate once QCD becomes strongly coupled, yielding a breaking of the electroweak group to electromagnetism at the QCD scale. If we imagine a parallel universe where the Higgs does not take a vev (or it is absent), below the QCD scale we would only see leptons generating anomalies of the EW symmetry which are canceled only when the quark sector is taken into account. We extrapolate this mechanism to beyond-the-SM physics.

This paper is structured as follows: in \cref{sec:Idea} we explore the deconstruction of two toy models that encode the main ideas and show how anomaly-cancellation is described both in an effective chiral description and a holographic description of the strong dynamics. The application of these ideas to specific SM deconstructions are discussed in \cref{sec:Models}. As a prototypical example, we present a novel 4321 implementation where the top quark is uniquely identified by the gauge symmetry, as justified by the smallness of the bottom and all other Yukawa couplings. We further discuss alternative SM deconstructions relaying on this mechanism. We conclude in \cref{sec:Conclusions}.

\section{General idea}
\label{sec:Idea}

Before tackling specific SM extensions, we present in this section two toy model examples that encapsulate, in a simpler manner, how we can use a composite sector for flavor deconstructions of a gauge symmetry.

\subsection{Toy model I}

Let us consider the deconstruction of an $SU(N)$ gauge theory with several families of a fermion field $\psi_{L,R}^i\;(i=1,\dots,N_\psi)$ transforming in a complex vector-like representation. The vector-like character of the theory makes it trivially anomaly-free. Let us assume we want to extend the original symmetry to $SU(N)_1\times SU(N)_2$. For fields charged under $SU(N)_{1(2)}$, we will say they belong to the first (second) site. The breaking of this extended gauge group to the diagonal $SU(N)_V$ can be triggered by a composite sector with a gauge symmetry $SU(N_{HC})$ and $N$ hyper-quarks $\zeta_{L,R}$ transforming in a vector-like complex representation. This sector has then a global symmetry $SU(N)_L\times SU(N)_R\times U(1)_V$, so the $SU(N)$ factors can be identified with the extended gauge symmetry, $SU(N)_{1(2)}\equiv SU(N)_{L(R)}$. When the hyper-sector confines, the hyper-quarks form a condensate, $\langle \bar\zeta_L^\alpha \zeta_R^{\beta} \rangle\propto \delta^{\alpha\beta}$, that breaks the symmetry to its diagonal subgroup:
\begin{align}\label{eq:SymBrPat}
SU(N)_L\times SU(N)_R\times U(1)_V\to SU(N)_V\times U(1)_V\,, 
\end{align}
so below the scale of this condensate, we recover the starting $SU(N)$ model. This breaking yields $N^2-1$ would-be-Goldstone bosons, all of them eaten by the gauge bosons that get mass.

In what follows, we assume that all complex representations are the fundamental, which we denote as $\square$. For the $\psi$ fermions, this is a natural choice as we intend to identify them with the SM fermions, as for the hyper-quarks we do so for simplicity. We can charge both chiralities of $\psi$ in the fundamental representation of the same $SU(N)_X$ (with $X=L,R$), or charge each of them under a different $SU(N)_X$ group. In either case, the theory potentially contains gauge anomalies that need to be canceled by introducing new fields. We start considering $N \geq 3$. Since all our fields are in fundamental representations of $SU(N)_{R}$ and $SU(N)_{L}$ respectively, they generate local gauge anomalies of the type $SU(N)_X-SU(N)_X-SU(N)_X$. To cancel them, we will add $N_{\chi}$ right-handed fermions, $\chi_R$, in the fundamental of $SU(N)_L$, and  $N_{\chi}$ left-handed fermions, $\chi_L$, in the fundamental of $SU(N)_R$, all of them singlets of $SU(N_{\rm HC})$. As we show below, these fermions can get a mass in the broken phase and get integrated out, leaving the same low-energy spectra that our initial model had.  The exact value of $N_{\chi}$ depends on the charge assignment of the $\psi$ fields. An example is illustrated in~\cref{tab:ToyModelI}, where we charge $\psi_L$ under $SU(N)_L$ and $\psi_R$ under $SU(N)_R$. If we have $N_{\chi}=N_{\rm HC}+N_\psi$, all gauge anomalies are canceled. For $N = 2$, there are no local gauge anomalies, but there could be global ones~\cite{Davighi:2019rcd} (also called non-perturbative anomalies). In particular, for every $SU(2)$ component, there should be an even number of fields charged in the fundamental representation~\cite{Witten:1982fp}. If not, the partition function will flip sign under large gauge transformations. If the number of fundamental fields on each site is even, we do not need $\chi_{L,R}$ fermions. Otherwise, an odd number of fields $\chi_{L,R}$ will be needed. Thus, for \cref{tab:ToyModelI}, we would only need one fermion $\chi_{L,R}$ if $N_{\rm HC}$ is even and none if $N_{\rm HC}$ is odd.

\begin{table}[t]
\begin{center}
\renewcommand{\arraystretch}{1.2}
\begin{tabular}{|c|c|c|c|}
\hline
Field & $SU(N_{\rm HC})$ & $SU(N)_L$ & $SU(N)_R$\\
\hline
\hline
$\zeta_L$ & $\square$ & $\square$ & $\mathbf{1}$\\
$\zeta_R$ & $\square$ & $\mathbf{1}$ & $\square$ \\
\hline
$\psi_L^i$ & $\mathbf{1}$ & $\square$ & $\mathbf{1}$ \\
$\psi_R^i$ & $\mathbf{1}$ & $\mathbf{1}$ & $\square$ \\
\hline
$\chi_L^a$ & $\mathbf{1}$ & $\mathbf{1}$ & $\square$ \\
$\chi_R^a$ & $\mathbf{1}$ & $\square$ & $\mathbf{1}$ \\
\hline
\end{tabular}
\end{center}
\caption{Toy model I: to cancel gauge anomalies, we need $N_{\chi}=N_{\rm HC}+N_\psi$, where $N_\psi$ and $N_{\chi}$ are the number of $\psi_{L,R}$ and $\chi_{L,R}$ fermions, respectively. Here, $i=1,\dots,N_\psi$ and $a=1,\dots,N_\chi$.}
\label{tab:ToyModelI}
\end{table}

Finally, to give mass to these new fermions, we assume the existence of an extended hyper-color sector that generates the four-fermion operators
\begin{align}\label{eq:4FMass}
\mathcal{L}_{\rm EHC} \supset \frac{\lambda_{ab}}{\Lambda_{\rm EHC}^2} (\bar \chi^a_L \chi^b_R)
(\bar \zeta_L \zeta_R)\,,
\end{align}
where $a$ is the flavor index of $\chi_{L,R}$ running from $1$ to $N_{\chi}$. When the chiral symmetry is broken, this interaction generates the mass terms
\begin{align}
\mathcal{L}_{\rm chiral} \supsetsim \frac{4\pi f_{\zeta}^3}{\Lambda_{\rm EHC}^2} \lambda_{ab}(\bar \chi^a_L \chi^b_R)\,,
\end{align}
so $\chi_{L,R}$ become vector-like fermions under the $SU(N)_V$ gauge symmetry.

\subsection{Toy model II}

Let us consider now as starting point an IR model with a gauge symmetry $SU(N) \times U(1)$ and multiple fermion families in the complex vector representation $\psi_{L,R}^i\sim (\square, 1)$ (with $i=1,\dots,N_\psi$). As before, this model is anomaly-free due to its vector-like character.

We can now deconstruct the $SU(N)$-factor of the gauge symmetry while keeping the $U(1)$ factor universal, so we extend the gauge symmetry to $SU(N)_1\times SU(N)_2\times U(1)$. The composite sector can be chosen in the same way as before: we assume an hyper-sector with gauge symmetry $SU(N_{HC})$ and $N$ hyper-quarks $\zeta_{L,R}$ triggering the same symmetry breaking described in~\cref{eq:SymBrPat}. In this case, there are several kinds of anomalies to consider:
\begin{itemize}
\item[(1)] {\it Cubic $U(1)$ anomalies, $U(1)-U(1)-U(1)$}. They cancel if
\begin{equation}
\sum_{\psi} (-1)^{s_\psi} (q_{\psi})^3 = 0,
\end{equation}
where $\psi$ denotes all fermions charged under $U(1)$ with charge $q_{\psi}$ and $s_{\psi}=1(0)$ for LH (RH) fermions.
This anomaly trivially cancels because $U(1)$ is universal and it cancels in the IR model.
\item[(2)] {\it Gravitational $U(1)$ anomalies.} They cancel if
\begin{equation}
\sum_{\psi} (-1)^{s_\psi} q_{\psi} = 0.
\end{equation}
Like cubic $U(1)$ anomalies, this anomaly cancels because it cancels in the IR model.
\item[(3)] {\it Pure $SU(N)_{i}$ anomalies, both local $SU(N)_X-SU(N)_X-SU(N)_X$ and global ones}. They can be canceled by introducing new fermions $\chi_{L,R}$ as in toy model I.
\item[(4)] {\it  Mixed anomalies $SU(N)_X-SU(N)_X-U(1)$}. They cancel if
\begin{equation}
\sum_{\psi} (-1)^{s_\psi} q_{\psi} = 0,\label{eq:MixedAnomalies}
\end{equation}
where the sum is restricted now to all the fermions charged under the same $SU(N)$ (that we assume that transform in the fundamental representation).
In principle, they only cancel if both $\psi_L$ and $\psi_R$ are located in the same site.
\end{itemize}

\begin{table}[t]
\begin{center}
\renewcommand{\arraystretch}{1.2}
\begin{tabular}{|c|c|c|c|c|c|}
\hline
Field & $SU(N_{\rm HC})$ & $SU(N)_1$ & $SU(N)_2$ & $U(1)$\\
\hline
\hline
$\zeta_L$ & $\square$ & $\square$ & ${\bf 1}$ & $-\frac{N_{\psi}}{N_{\rm HC}}$ \\
$\zeta_R$ & $\square$ & ${\bf 1}$ & $\square$ & $-\frac{N_{\psi}}{N_{\rm HC}}$  \\
\hline
$\psi_L^i$ & ${\bf 1}$ & $\square$ & ${\bf 1}$ & $1$ \\
$\psi_R^i$ & ${\bf 1}$ & ${\bf 1}$ & $\square$ & $1$  \\
\hline
$\chi_L^a$ & $\mathbf{1}$ & $\mathbf{1}$ & $\square$ & $0$ \\
$\chi_R^a$ & $\mathbf{1}$ & $\square$ & $\mathbf{1}$ & $0$ \\
\hline
\end{tabular}

\bigskip
\renewcommand{\arraystretch}{1.2}
\begin{tabular}{|c|c|c|c|c|}
\hline
Field & $SU(N_{\rm HC}+N_\psi)$ & $SU(N)_1$ & $SU(N)_2$ \\
\hline
\hline
$\Xi_L=(\zeta_L,\psi_L^i)$ & $\square$ & $\square$ & ${\bf 1}$ \\
$\Xi_R=(\zeta_R,\psi_R^i)$ & $\square$ & ${\bf 1}$ & $\square$ \\
\hline
$\chi_L^a$ & $\mathbf{1}$ & $\mathbf{1}$ & $\square$ \\
$\chi_R^a$ & $\mathbf{1}$ & $\square$ & $\mathbf{1}$ \\
\hline
\end{tabular}
\end{center}
\caption{Toy model II (top) and its possible UV-completion (bottom). To cancel gauge anomalies, we need $N_{\chi}=N_{\rm HC}+N_\psi$, where $N_\psi$ and $N_{\chi}$ are the number of $\psi_{L,R}$ and $\chi_{L,R}$ fermions, respectively. Here, $i=1,\dots,N_\psi$ and $a=1,\dots,N_\chi$.}
\label{tab:ToyModelII}
\end{table}

The last type of anomaly appears to be an obstacle to put $\psi_L$ and $\psi_R$ on different sites, not fixable by introducing new degrees of freedom.
However, as shown in the example given in~\cref{tab:ToyModelII} (top), these anomalies can be canceled by appropriately charging the hyper-quarks under the $U(1)$ symmetry. Hyper-quarks then also contribute to the sums of \cref{eq:MixedAnomalies} and make them vanish without creating other $U(1)$ anomalies due to their vector-like character under $U(1)$.

Interestingly, this solution has a very natural UV completion that avoids fractional charges: we can promote $SU(N_{\rm HC})$ to $SU(N_{\rm HC}+N_\psi)$ and arrange the fields as shown in~\cref{tab:ToyModelII} (bottom). Then, if $SU(N_{\rm HC}+N_\psi)$ breaks to $ SU(N_{\rm HC})\times U(1)$ at some high scale, we recover the original model. This approach is similar to the Pati-Salam (PS) unification of quarks and leptons~\cite{Pati:1974yy} but now applied to the hyper-sector. In this case, the $\psi$ fields become the hyper-leptons of the hyper-quarks.

\subsection{Effective chiral description}
\label{sec:EffChi}

When the composite sector confines, its degrees of freedom change from hyper-colored states to Goldstone bosons and resonances. We can describe our models using these degrees of freedom in the same way chiral perturbation theory ($\chi$PT) describes the low-energy limit of QCD. In such an effective chiral description, non-hyper-colored fermions generate anomaly contributions that are only canceled by the so-called Wess-Zumino-Witten (WZW) terms in the effective Lagrangian~\cite{Wess:1971yu,Yonekura:2020upo,Lee:2020ojw}, constructed with the (would-be) Goldstone bosons resulting from the breaking. In this subsection, we explicitly build such terms for our toy models and the models discussed in the next section. This subsection and \cref{sec:Hologr} are more technical than the rest of the article, and they could be skipped in a first reading.

In this subsection, we will work in Euclidean time and a compactified 4d spacetime, $\mathcal{M}_4=S^4$. 
For field configurations on this spacetime that can be continuously deformed to the constant map, the easiest way to write WZW terms is to follow Witten's construction in~\cite{Witten:1983tw}. WZW terms are the integral of a 5-form $\omega^{\rm WZW}_5$ in a fictitious $5$d disk $\mathcal{M}_5$ whose boundary is the physical $4$d space $\mathcal{M}_4=\partial \mathcal{M}_5$,
\begin{equation}
S_{\rm WZW} = \int_{\mathcal{M}_5} \omega^{\rm WZW}_5.
\end{equation}
The form $\omega^{\rm WZW}_5$ has to be invariant under the symmetries of the theory and closed when it is extended to a higher-dimensional space with $d>5$, ${\rm d}\omega^{\rm WZW}_5 = 0$.\footnote{In components, $({\rm d}\omega)_{\mu_0\dots \mu_{p}}\equiv(p+1)\,\partial_{[\mu_0}\omega_{\mu_1\dots \mu_{p}]}$.}
Then, this term can be written as the integral on the physical 4d space $\mathcal{M}_4$ of the primitive 4-form $\omega^{\rm WZW}_4$,  ${\rm d}\omega^{\rm WZW}_4 = \omega^{\rm WZW}_5$, which does not change under continuous deformations of the 5d extension.
If several classes of 5d extensions 
that are not connected by a continuous deformation exist, they may give different values of $S_{\rm WZW}$.
A quantization condition for the normalization of $\omega_5^{\rm WZW}$ is then necessary~\cite{Witten:1983tw,Alvarez:1984es} to ensure that different 5d extensions give the same value of $e^{i S_{\rm WZW}}$, which is the physically relevant object.
Thus, the particular choice to extend the fields to the fictitious 5d space is irrelevant, and therefore, unphysical. However, it is typically easier to write $\omega^{\rm WZW}_5$ which, contrary to $\omega^{\rm WZW}_4$, explicitly preserves the symmetries of the theory.\footnote{An extra more technical assumption on $\omega^{\rm WZW}_5$ called the {\it Manton condition} is necessary to properly define $G$-invariant WZW terms in a coset space, $G/H$, with $H$ a closed subgroup of $G$~\cite{Davighi:2018inx,Davighi:2020vcm}. 
This condition is automatically satisfied if 
the 4-th homology group of the coset space is trivial, $H_4(G/H)=0$, or if $G$ is semisimple. 
For the coset spaces we consider here, $G/H \cong SU(N)$ and $H_4(SU(N))=0$.}

Before building the WZW terms, it is convenient to introduce the Chern-Simons (CS) forms $\omega_{2n+1}^{\rm CS}$ and some of their properties~\cite{Weinberg:1996kr}. Let $A\equiv A^a_{\mu} T_a {\rm d} x^{\mu}$ be a matrix-valued gauge connection that, for our purposes, will be a connection of $SU(N)$. We assume the gauge fields here are normalized such that the kinetic term comes with the inverse squared of the gauge coupling constant. CS terms are defined as $2n+1$-forms such that
\begin{align}
{\rm d} \omega_{2n+1}^{\rm CS} = I_{2n+2}(A)\,,
\end{align}
with 
\begin{align}
I_{2n}(A)={\rm Tr}(\underbrace{F\wedge \ldots \wedge F}_{n~{\rm times}})\,.
\end{align}
CS forms can be explicitly built like
\begin{align}
\omega_{2n+1}^{\rm CS}(A) = (n+1)\int_0^1 {\rm d}t\,
{\rm Tr} \big(A \wedge\underbrace{F_t\wedge \ldots \wedge F_t}_{n~{\rm times}}\big)\,,
\end{align}
where $F_t$ is the field-strength tensor of the gauge connection $A_t=t A$. 
Although the CS term is not invariant under gauge transformations, its variation under the infinitesimal gauge transformation $\alpha$ is the differential of a $2n$ form,
\begin{align}\label{eq:omega12n}
\delta_{\alpha} \omega_{2n+1}^{\rm CS}(A)={\rm d} \omega^1_{2n}(\alpha,A)\,,
\end{align}
with
\begin{align}
\omega^1_{2n}(\alpha,A)=n(n+1)\int_0^1 {\rm d}t\, (1-t){\rm Tr}\{ \alpha\, {\rm d}(A \wedge\underbrace{F_t\wedge \ldots \wedge F_t}_{n-1~{\rm times}}) \}\,.
\end{align}
Interestingly, $\omega^1_{2n}$ describes possible anomalies from fermion loops. Indeed, in the case of toy models I and II, non-hyper-colored fermion loops contribute to the $SU(N)_X^3$ anomaly as
\begin{align}\label{eq:SUN3Anom}
\delta_{\alpha} \Gamma \supset \frac{i(N_{\psi}-N_{\chi})}{24 \pi^2}\int_{\mathcal{M}_4 }[\omega^1_4(\alpha_L,A_L)-\omega^1_4(\alpha_R,A_R)+\delta_{\alpha} B_4(A_L,A_R)]\,,
\end{align}
where $\alpha=(\alpha_L,\alpha_R)$, $\Gamma$ is the quantum effective action, and $B_4$ are local counterterms that one may add to shift the anomaly between the different currents~\cite{Chu:1996fr} and depend on the regularization procedure of the loop integrals~\cite{Weinberg:1996kr}. We will assume here that the currents associated to all generators are treated symmetrically, and therefore, $B_4=0$. Note that for $SU(2)$, $\omega_5^{\rm CS}=0$ because the $SU(2)$ generators satisfy ${\rm Tr}(T_a\{T_b,T_c\})=0$, showing that, indeed, $SU(2)$ cannot have pure local anomalies.

For the mixed anomaly of toy model II, it is also convenient to introduce the {\it mixed} CS term, satisfying in this case
\begin{align}\label{eq:mixedCS}
{\rm d}\bar \omega_5^{\rm CS}(A,A_{U(1)})=F_{{U(1)}}\wedge I_4(A)\,,
\end{align}
where $A_{U(1)}$ and $F_{U(1)}$ are the gauge field and field strength tensor of the extra $U(1)$. They read
\begin{align}
\bar \omega_5^{\rm CS}(A,A_{U(1)})=\, \xi\,F_{U(1)}\wedge\omega^{\rm CS}_3(A)
+(1-\xi) A_{U(1)}\wedge I_4(A)\,,
\end{align}
with $\xi$ parametrizing an exact form we may add without affecting~\cref{eq:mixedCS}. Similarly to the pure CS form, its variation is an exact form, $\delta_{(\alpha,\alpha_{U(1)})} \bar \omega_{5}^{\rm CS}
={\rm d} \bar \omega^1_{4}$, such that
\begin{align}\label{eq:baromega12n}
\bar \omega^1_{4}(\alpha,\alpha_{U(1)};A,A_{U(1)})=\xi F_{U(1)}\wedge\omega^1_2(\alpha,A)+
\alpha_{U(1)} (1-\xi) I_4(A)\,,
\end{align}
with $\alpha_{U(1)}$ the gauge parameter of the $U(1)$ group. Thus, the $N_{\psi}$ fermions $\psi_{L,R}$ with charge $1$ under $U(1)$ in toy model II contribute to the mixed anomalies $SU(N)_X-SU(N)_X-U(1)$ as
\begin{align}\label{eq:SUN2U1Anom}
\delta_{\alpha} \Gamma \supset \,&\frac{iN_{\psi}}{32 \pi^2}\int_{\mathcal{M}_4 }\left[
\bar \omega^1_4(\alpha_L,\alpha_{U(1)};A_L,A_{U(1)})-\bar \omega^1_4(\alpha_R,\alpha_{U(1)};A_R,A_{U(1)})
\right]
\nonumber\\
=&\frac{iN_{\psi}}{32 \pi^2}\int_{\mathcal{M}_4 }\left\{\xi F_{U(1)}\wedge[\omega^1_2(\alpha_L,A_L)-\omega^1_2(\alpha_R,A_R)]+
\alpha_{U(1)} (1-\xi) [I_4(A_L)-I_4(A_R)]
\right\}\,,
\end{align}
where now $\alpha=(\alpha_L,\alpha_R,\alpha_{U(1)})$, and $\xi$ parametrizes how the anomaly is shared between the $U(1)$ current and $SU(N)$ currents. For instance, $\xi=0\,(1)$ shifts the anomaly completely to the $U(1)$ current ($SU(N)$ currents).

To construct the WZW terms, we follow~\cite{Chu:1996fr}, where the authors give the general construction for a broad class of $G/H$ coset spaces. We particularize here for the case $SU(N)\times SU(N)/SU(N) \cong SU(N)$.
First, it is convenient to define the form
\begin{align}
\tilde \omega_{2n+1} (A_0,A_1) = (n+1)\int_0^1 {\rm d}t\,
{\rm Tr} \big((A _1-A_0)\wedge\underbrace{F_t\wedge \ldots \wedge F_t}_{n~{\rm times}}\big)\,,
\end{align}
where here $F_t$ is the field strength tensor of the gauge connection $A_t=(1-t)A_0 + t A_1$. This form is invariant under simultaneous gauge transformations of $A_0$ and $A_1$, $A^g= g(A+{\rm d})g^{-1}$,
\begin{align}\label{eq:wGInv}
\tilde \omega_{2n+1} (A^g_0,A^g_1) = \tilde \omega_{2n+1} (A_0,A_1)\,,
\end{align}
and it satisfies~\cite{Chu:1996fr}
\begin{align}\label{eq:dtildew}
{\rm d} \tilde \omega_{2n+1} (A_0,A_1) = I_{2n+2}(A_1)-I_{2n+2}(A_0)\,.
\end{align}
Note also that $\omega^{\rm CS}_{2n+1} (A)=\tilde \omega_{2n+1} (0,A)$.

We first consider the case of $SU(N)$ with $N\geq 3$.\footnote{The 4-th homotopy group of the coset space is trivial, $\pi_4(SU(N))=0$ for $N\geq 3$, so every configuration of the Goldstone field can be extended to $\mathcal{M}_5$.}
The WZW terms are functions of the gauge fields and the would-be Goldstone bosons. The latter will be written as a matrix $\Sigma \in SU(N)$, transforming under the total group $SU(N)_L\times SU(N)_R$ as $\Sigma\to U_L \,\Sigma \,U_R^{\dagger}$. In the expressions below, this matrix will be written like $\Sigma= g_L^{-1} g_R$ for some choice of $g_L$ and $g_R$.
A first candidate for the WZW term, up to the appropriate normalization, would be
\begin{align}\label{eq:omegaprima3}
\left.\omega_5^{\prime}\left(\Sigma,A_L,A_R\right) \right|_{SU(N)^3}=&\,
-\tilde \omega_5 \left( \frac{A^{g_L}_L+ A^{g_R}_R}{2} ,A^{g_L}_L\right) + 
\tilde \omega_5 \left(   \frac{A^{g_L}_L+ A^{g_R}_R}{2} ,A^{g_R}_R\right)\,.
\end{align}
The particular way $\Sigma$ is split into $g_L$ and $g_R$ is irrelevant, and all of them give the same result due to \cref{eq:wGInv} (a possible choice is, for instance, $g_L=\mathbb{1}$ and $g_R=\Sigma$). It is easy to check that this form is invariant under gauge transformations, and using \cref{eq:dtildew}, that
\begin{align}
{\rm d} \omega_5^{\prime} = -I_6(A_L)+ I_6(A_R).
\end{align}
This form is closed if only the unbroken group is gauged so $A_L=A_R$.
We would like however to gauge the full group. For that, we will add the appropriate CS terms, defining our $SU(N)^3$ WZW term to be~\cite{Hull:1990ms,Brauner:2018zwr}:
\begin{align}\label{eq:WZWSUN3}
\left.\omega_5^{\rm WZW}\left(\Sigma,A_L,A_R\right) \right|_{SU(N)^3}=&\,\frac{m}{24 \pi^2}\Bigg[
\left.\omega_5^{\prime}\left(\Sigma,A_L,A_R\right) \right|_{SU(N)^3}
+ \omega^{\rm CS}_5(A_L)
- \omega^{\rm CS}_5(A_R)\Bigg]\,,
\end{align}
where, as discussed above, the overall normalization has to be quantized, $m\in \mathbb{Z}$.\footnote{The 5-th homotopy group of the coset space is non-trivial, $\pi_5(SU(N))=\mathbb{Z}$ for $N\geq 3$, which implies that there are infinite classes of 5d extensions for $\Sigma$ not connected by a deformation.}
Due to the CS forms, the WZW term is closed but invariant only under gauge transformations in the unbroken group $SU(N)_V$. The variation under a general gauge transformation is proportional to the integral of $ {\rm d}\omega^1_4(\alpha_L,A_L)- {\rm d}\omega^1_4(\alpha_R,A_R)$ in the 5d space. Using the Gauss theorem, it can be written localized in the boundary, i.e. in the physical space. If $m=N_{\chi}-N_{\psi}=N_{HC}$, this contribution precisely cancels the $SU(N)_X^3$ anomaly caused by the non-hyper-colored fermions of \cref{eq:SUN3Anom}.

For $SU(N)$ groups with $N=2$, the $SU(2)^3$ WZW term vanishes. However, if $N_{\rm HC}=N_{\chi}-N_{\psi}$ is odd, the chiral description of toy models I and II has a non-perturbative WZW-like term (different to the ones discussed here) that reproduces the global-anomaly contribution from the hyper-quarks and cancels the contribution from the non-hyper-colored fermions. 
This is achieved if this term flips the sign of the partition function for configurations of the $\Sigma$ field that cannot be deformed to the constant map.\footnote{Note that the 4-th homotopy group of the coset space is $\pi_4(SU(2))=\mathbb{Z}_2$. A large gauge transformation of one of the $SU(2)$ factors changes the homotopy class of the field configuration of $\Sigma$.}

Coming back to general values of $N\geq 2$, in toy model II there is also a $U(1)$ gauge field that allows to define a new form~\cite{Brauner:2018zwr}
\begin{align}
\left.\omega^{\prime}_5\left(\Sigma,A\right)\right|_{SU(N)^2-U(1)} =\,F_{U(1)}\wedge\left[
-\tilde \omega_3 \left( \frac{A^{g_L}_L+ A^{g_R}_R}{2} ,A^{g_L}_L\right) + 
\tilde \omega_3 \left( \frac{A^{g_L}_L+ A^{g_R}_R}{2} ,A^{g_R}_R\right)\right]\,.
\end{align}
As before, this form is gauge invariant, but not closed under general gauge transformations:
\begin{align}
{\rm d}\omega^{\prime}_5\left(\Sigma,A\right)\big|_{SU(N)^2-U(1)} = F_{U(1)}\wedge [-I_4(A_L)+I_4(A_R)]\,.
\end{align}
To make it closed, we add mixed CS forms and define the $SU(N)^2-U(1)$ WZW term:
\begin{align}
\left.\omega^{\rm WZW}_5\left(\Sigma,A\right)\right|_{SU(N)^2-U(1)} =\frac{\bar m}{32\pi^2}
\bigg\{\,
&\left.\omega^{\prime}_5\left(\Sigma,A\right)\right|_{SU(N)^2-U(1)}
\nonumber \\
&+\xi\, F_{U(1)}\wedge\left[\omega^{\rm CS}_3(A_L)
- \omega^{\rm CS}_3(A_R)\right] \nonumber \\
&+(1-\xi)\, A_{U(1)}\wedge\left[I_4(A_L)
-I_4(A_R)\right]
\bigg\}.\label{eq:WZWSUN2U1}
\end{align}
Like in the previous case, the WZW term is now only invariant under the unbroken group but not under the complete group. The normalization here can be chosen for the variation of the CS terms to exactly cancel the anomaly contribution from the $N_{\psi}$ fermions $\psi^i_{L,R}$ of~\cref{eq:SUN2U1Anom}, $\bar m = -N_{\psi}$. 
Thus, this contribution makes the effective theory anomaly free. Remarkably, in this case we can directly write a primitive 4-form ${\rm d} \omega^{\rm WZW}_4 = \omega^{\rm WZW}_5$, avoiding fictitious 5d extensions:
\begin{align}
\left.\omega_4^{\rm WZW}\right|_{SU(N)^2-U(1)} =
\frac{\bar m}{32\pi^2}\,A_{U(1)}\wedge & \bigg\{
-\tilde \omega_3 \left( \frac{A^{g_L}_L+ A^{g_R}_R}{2} ,A^{g_L}_L\right) + 
\tilde \omega_3 \left(   \frac{A^{g_L}_L+ A^{g_R}_R}{2} ,A^{g_R}_R\right)\nonumber\\
&+\xi\left[\omega_3^{\rm CS} (A_L)-\omega_3^{\rm CS} (A_R)\right]
\bigg\}.\label{eq:WZWSUN2U14d}
\end{align}

\subsection{Holographic realization}
\label{sec:Hologr}

Holographic models are a useful way to describe strongly-coupled sectors in the large-$N_{\rm HC}$ limit. Due to holography or the AdS/CFT correspondence~\cite{Maldacena:1997re,Witten:1998qj}, one expects that theories living in a slice of a 5d space with metric
\begin{align}
{\rm d} s^2 = e^{-2 \sigma(y)}\eta_{\mu\nu} 
{\rm d} x^{\mu}
{\rm d} x^{\nu} - {\rm d} y^2,\label{eq:AdS5}
\end{align}
where $\sigma(y) \sim y $ when $y \to -\infty$,
describe effectively strongly coupled sectors in 4d. This 5d space is an asymptotic AdS space, and it will be bounded by a 4d UV brane at $y=y_{\rm UV}$ and a 4d IR brane at $y=y_{\rm IR}>y_{\rm UV}$. In this description, the hyper-colored degrees of freedom are replaced by this 5d bulk with weakly-coupled physics. It is instructive to see how this 5d dynamics provides mechanisms to cancel the anomaly contribution of the non-hyper-colored fermions.

The holographic dictionary~\cite{Gherghetta:2010cj} guides us in how to map the features of the strongly-coupled sector into the 5d model. We can then build the holographic version of toy models I and II. In this subsection, we only consider models with $SU(N)$ groups with $N\geq 3$.
They consist on a $SU(N)_L\times SU(N)_R\times U(1)_V$ gauge theory in the 5d bulk corresponding to a composite sector with a $SU(N)_L\times SU(N)_R\times U(1)_V$ global symmetry.\footnote{So far, we have not considered the axial $U(1)_A$ symmetry because it is anomalous. However, in the large-$N_{\rm HC}$ limit this anomaly is subleading so $U(1)_A$ could also be included. Nevertheless, it does not play any relevant role in our discussion.}
Reproducing the anomalies of the symmetries of the composite sector requires the inclusion of CS terms in the bulk that we discuss below~\cite{Witten:1998qj,Hill:2006wu}.
Let us focus first on the $SU(N)_L\times SU(N)_R$ part, which is common for both toy models.
A condensate breaking $SU(N)_L\times SU(N)_R$ to the diagonal $SU(N)_V$ can be implemented through the breaking of the symmetry in the IR brane. This is done by selecting as boundary conditions of the gauge fields those that set to zero the 4d components associated to the broken generators, so $SU(N)_V$ is preserved in the IR brane. The UV brane will preserve the full $SU(N)_L\times SU(N)_R$ symmetry because it is gauge in the dual 4d model.
The fundamental degrees of freedom $\chi_{L,R}$ and $\psi_{L,R}$ appear as fermions localized in the UV brane. In principle, they will create an anomaly localized in the UV brane similar to the one in~\cref{eq:SUN3Anom},
\begin{align}
\delta_{\alpha}\Gamma_{\rm 5d} \supset \delta(y-y_{\rm UV})\frac{i(N_{\psi}-N_{\chi})}{24 \pi^2}\left[ \omega_4^1(\alpha_L,A_L)-\omega_4^1(\alpha_R,A_R)\right].\label{eq:UVAnomaly}
\end{align}
This anomaly is canceled by the CS terms in the bulk~\cite{Arkani-Hamed:2001uol,Gripaios:2007tk}:
\begin{align}
\mathcal{L}_{5d} \supset \frac{N_{\rm HC}}{24\pi^2} \left[\omega_5^{\rm CS}(A_L)-\omega_5^{\rm CS}(A_R)\right].\label{eq:HoloCS}
\end{align}
The variation of these terms exactly cancels the anomalous contribution of \cref{eq:UVAnomaly}, and does not create a new one in the IR brane because, by boundary conditions, $A_L(y_{\rm IR})=A_R(y_{\rm IR})$.

Concerning $U(1)_V\equiv U(1)$
in the holographic toy model I, it is preserved in the IR brane because it is respected by the condensate, but broken in the UV brane by boundary conditions $A_{U(1)}|_{y_{\rm UV}}=0$ on the 4d components, because $U(1)$ is not gauged in the dual theory.
However, in the holographic toy model II, the $U(1)$ symmetry is preserved everywhere, including the UV brane. In this case, the mixed anomaly generated in the UV brane similar to~\cref{eq:SUN2U1Anom} is canceled by the mixed CS term in the bulk,
\begin{align}
\mathcal{L}_{5d} \supset \frac{\bar m}{32\pi^2}\left\{ \xi\, F_{U(1)}\wedge\left[\omega^{\rm CS}_3(A_L)
- \omega^{\rm CS}_3(A_R)\right] 
+(1-\xi) A_{U(1)}\wedge\left[I_4(A_L)
- I_4(A_R)\right]\right\},\label{eq:HoloMixCS}
\end{align}
with $\bar m = -N_{\psi}$ and which, as before, does not generate any anomaly in the IR brane because $A_L(y_{\rm IR})=A_R(y_{\rm IR})$.\footnote{Although irrelevant for gauge-anomaly cancellation, a similar CS term is also present in the holographic toy model I to match the $SU(N)-SU(N)-U(1)_V$ 't Hooft anomalies of the composite sector.}

The cancellation of these anomalies is reminiscent of the cancellation of the same anomalies in the effective chiral description with WZW terms. In both cases, the same CS terms are the ones that cancel anomalies from the non-hyper-colored fermions, in one case in the AdS space, and in the other in a fictitious 5d space. The only condition to be able to write such terms is that CS forms must vanish when evaluated in the unbroken gauge fields. In the AdS space description this condition is required to not create anomalies in the IR brane, while in the effective chiral description this is needed to build the WZW term. This is not surprising as it is known that CS terms in the 5d bulk are dual to WZW terms in the 4d description. Indeed, one can show that after integrating out the bulk degrees of freedom of a 5d theory, CS terms in the 5d action generate WZW terms in the so-called holographic action~\cite{Panico:2007qd}.

\section{Deconstructing the Standard Model}
\label{sec:Models}

We now apply the ideas from the previous section to build realistic models based on SM deconstructions. Before doing so, it is useful to embed the SM gauge group into larger global symmetries present in the SM kinetic terms, which are anomaly free. For instance, $U(1)_Y=U(1)_R \oplus U(1)_{B-L}$, where $U(1)_R$ charges with $\pm 1/2$ the up (down) fermions, and $U(1)_{B-L}$ charges with $-1/2\, (1/6)$ for leptons (quarks). Furthermore, $U(1)_R\subset SU(2)_R$ and $SU(3)_c\times U(1)_{B-L} \subset SU(4)_{PS}$. We thus arrive to the PS symmetry~\cite{Pati:1974yy}: if we add one singlet fermion per family, i.e. three right-handed neutrinos $\nu_R$, each SM family can be embedded into two multiplets $\Psi_L \sim ({\bf 4},{\bf 2},{\bf 1})$ and $\Psi_R \sim ({\bf 4},{\bf 1},{\bf 2})$ of the PS group $SU(4)_{PS}\times SU(2)_L\times SU(2)_R$:
\begin{align}
\Psi_L=
\begin{pmatrix}
q_L \\
\ell_L
\end{pmatrix}
\,,\qquad
\Psi_R=
\begin{pmatrix}
(u_R && d_R) \\
(\nu_R && e_R)
\end{pmatrix}
\,.
\end{align}
One advantage of this embedding is that anomaly cancellation for each SM family becomes transparent: neither mixed nor gravitational anomalies can appear because the group is semisimple, $SU(4)_{PS}$ anomaly cancels between $\Psi_L$ and $\Psi_R$ (as it is a vector-like symmetry), and $SU(2)_{L}$ and $SU(2)_{R}$ are anomaly-safe groups. Furthermore, there are no global anomalies associated to $SU(2)_{L,R}$ because for each group there are 4 doublets per family forming a fundamental of $SU(4)_{PS}$. Another advantage of considering the PS group embedding compared to other unification groups, such as $SU(5)$ or $SO(10)$, is that the gauging of this symmetry does not introduce proton decay, thus allowing for typically lower (partial) unification scales.\footnote{See~\cite{FernandezNavarro:2023hrf} for a recent model example of $SU(5)$ deconstruction.} 

For our purposes, even when we do not gauge the PS group entirely, it will be convenient to think in terms of this symmetry. When deconstructing the SM gauge symmetry, we distinguish the following scenarios:
\begin{itemize}
    \item[i)] When full PS multiplets are present in a given site, no extra $U(1)$ symmetries are involved and the situation is similar to toy model I. In this case, we only need to appropriately choose the number of vector-like fermions $\chi_{L,R}$, if needed, to make the deconstruction anomaly-free.
    \item[ii)] When a PS multiplet is distributed among different sites and the only deconstructed groups are semisimple, a universal $U(1)$ symmetry, which can be hypercharge or some component of it, will cause mixed anomalies. In this situation, one can use the mechanisms of toy model~II for anomaly cancellation.
    \item[iii)] A last possibility appears when a PS multiplet is distributed among different sites and the deconstructed symmetry involves $U(1)$ factors. In this class of models, the mechanisms discussed in model I or II are not sufficient to cancel gauge anomalies, so we disregard it in what follows.
\end{itemize}
We provide phenomenologically relevant examples of SM deconstructions of type i) and ii) in the next subsections.

\subsection{4321 deconstructions}

In this section, we focus in deconstructions of the SM color group based on the gauge symmetry
\begin{align}
\mathcal{G}_{4321}\equiv SU(4)\times SU(3)^\prime\times SU(2)_L\times U(1)_X\,,
\end{align}
where $SU(2)_L$ acts universally on the three SM fermion families, while the $SU(4)$, $SU(3)^\prime$ and $U(1)_X$ groups act non-universally. Around the TeV scale, this gauge group is assumed to get spontaneously broken to the SM subgroup, such that $SU(3)_c\times U(1)_Y\equiv \left[SU(4)\times SU(3)^\prime\times U(1)_X\right]_{\rm diag}$ and $SU(2)_L$ is the SM group factor. Hence, gauge universality of the SM appears as an emergent property at low-scales for this class of models. Rather than discussing the fermion charges in terms of the 4321 gauge symmetry, it results more convenient to present them in terms of the larger (global) symmetry
\begin{align}
\mathcal{G}_{4422}\equiv SU(4)\times SU(4)^\prime\times SU(2)_L\times SU(2)_R\,,
\end{align}
with $SU(4)^\prime\supset SU(3)^\prime\times U(1)^\prime$ being non-universal on the SM families and, similarly to $SU(2)_L$, the $SU(2)_R\supset U(1)_R$ symmetry assumed to be universal. The $U(1)_X$ factor in the 4321 symmetry is obtained from the diagonal combination of the $U(1)$ factors in $SU(4)^\prime$ and $SU(2)_R$, namely $U(1)_X\equiv[U(1)^\prime\times U(1)_R]_\mathrm{diag}$.

In analogy with the toy model examples from \cref{sec:Idea}, we will say that fields charged under $SU(4)$ ($SU(4)^\prime$) belong to the first (second) site. Different charge assignments for the SM fermions correspond to different 4321 implementations. The most common one locates first- and second-family fermions in the second site by charging them as in the SM under $SU(3)^\prime\times SU(2)_L\times U(1)_X$, whereas third-generation quarks and leptons (together with a right-handed neutrino) are unified into $SU(4)$ fourplets,\footnote{For the usual see-saw mechanisms, the scale of quark-lepton unification has to be as around $10^{14}$ GeV to have light neutrinos without fine tuning. Instead, naturally light neutrino masses with a low $SU(4)$-breaking scale can be realized through an inverse see-saw mechanism by introducing an additional gauge-singlet fermion~\cite{FileviezPerez:2013zmv,Greljo:2018tuh,Fuentes-Martin:2020pww,Fuentes-Martin:2022xnb}.} thus belonging to the first site. The breaking to the SM group is typically done through a set of scalar fields charged under $SU(4)$ that acquire a vev. An alternative possibility consists in breaking the 4321 symmetry through the condensate of hyper-fermions from a strongly-coupled sector~\cite{Fuentes-Martin:2020bnh}.\footnote{In 4321 models, an extra advantage of breaking the symmetry with a condensate is that one avoids dangerous 5-dimensional operators that break baryon number, leading to fast proton decay even when they are suppressed by the Planck scale~\cite{Dutka:2022lug}. We thank Tomasz Dutka for pointing this out.}

\begin{table}[t]
\begin{center}
\renewcommand{\arraystretch}{1.2}
\begin{tabular}{|c|c|c|c|c|c|c|}
\hline
Field & $SU(N_{\rm HC})$ & $SU(4)$ & $SU(4)^\prime$ & $SU(2)_L$ & $SU(2)_{R}$\\
\hline
\hline
$\zeta_R$ & $\square$ & $\mathbf{4}$ & $\mathbf{1}$ & $\mathbf{1}$ & $\mathbf{1}$ \\
$\zeta_L$ & $\square$ & $\mathbf{1}$ & $\mathbf{4}$ & $\mathbf{1}$ & $\mathbf{1}$ \\
\hline
$\chi^i_L$ & $\mathbf{1}$ & $\mathbf{4}$ & $\mathbf{1}$ & $\mathbf{2}$ & $\mathbf{1}$  \\
$\chi^i_R$ & $\mathbf{1}$ & $\mathbf{1}$ & $\mathbf{4}$ & $\mathbf{2}$ & $\mathbf{1}$  \\
\hline
$\psi^3_L$ & $\mathbf{1}$ & $\mathbf{4}$ & $\mathbf{1}$ & $\mathbf{2}$ & $\mathbf{1}$ \\
$\psi^3_R$ & $\mathbf{1}$ & $\mathbf{4}$ & $\mathbf{1}$ & $\mathbf{1}$ & $\mathbf{2}$ \\
$\psi^{1,2}_L$ & $\mathbf{1}$ & $\mathbf{1}$ & $\mathbf{4}$ & $\mathbf{2}$ & $\mathbf{1}$ \\
$\psi^{1,2}_R$ & $\mathbf{1}$ & $\mathbf{1}$ & $\mathbf{4}$ & $\mathbf{1}$ & $\mathbf{2}$ \\
\hline
\end{tabular}
\end{center}
\caption{4321 deconstruction from~\cite{Fuentes-Martin:2020bnh}. To cancel the anomalies, we need $N_{\rm HC}=2N_{\chi}$.}
\label{tab:SU4Deconstruction_0}
\end{table}

The strongly-coupled option follows a similar structure to the one exemplified by toy model~I, with new will-be-vector-like fermions being the ones responsible for compensating the gauge anomalies introduced by the hyper-fermions (see \cref{tab:SU4Deconstruction_0}).\footnote{In this implementation, we choose $\zeta_R$ ($\zeta_L$) to be a fundamental of $SU(4)$ ($SU(4)^{\prime}$). In principle, the opposite identification is also possible, exchanging the chiralities of the will-be-vector-like fermions. Other representations of the will-be-vector-like fermions under $SU(2)_L$ or $SU(2)_R$ are also possible.} This choice of charges for the SM fermions singularizes the third family, resulting in a $U(2)_q\times U(2)_u\times U(2)_d\times U(2)_\ell\times U(2)_e$ flavor symmetry before 4321 symmetry breaking. Among other things, this symmetry provides an explanation for the smallness of the 2-3 CKM matrix elements and yields an extra protection for the NP sector against flavor constraints. Interestingly, if there is a mass term between $\chi_R^i$ and $q^{1,2}_L$, the will-be-vector-like fermions induce the Yukawa couplings which are a priori forbidden by the gauge symmetry when they are integrated out. Moreover, it becomes easy to embed this construction into a more complete setup where the Higgs is also localized in the first site, hence explaining the smallness of first- and second-generation Yukawas. However, the smallness of the bottom and tau Yukawas typically remains unexplained by this implementation, as all third-family fermions are put in the same footing.

Changing the number of will-be-vector-like fermions with respect to $N_{\rm HC}$ leads to different arrangements of the SM fields. If $N_{\rm HC}=2N_{\chi}+2$, anomalies would cancel if, for instance, third-family right-handed fields are charged under $SU(4)^{\prime}$. This possibility results in a $U(2)_q\times U(3)_u\times U(3)_d\times U(2)_\ell\times U(3)_e$ flavor symmetry before 4321 symmetry breaking, which is enough to explain the suppression of light Yukawas and 2-3 CKM matrix elements~\cite{Greljo:2023bix}. Indeed, if the SM Higgs is embedded in a scalar field charged under $\mathcal{G}_{4321}$ like $({\bf 4},{\bf \bar 3},{\bf 2})_{-\frac{1}{2}}$, only the top Yukawa coupling can be written at the renormalizable level. 

\begin{table}[t]
\begin{center}
\renewcommand{\arraystretch}{1.2}
\begin{tabular}{|c|c|c|c|c|c|c|}
\hline
Field & $SU(N_{\rm HC})$ & $SU(4)$ & $SU(4)^\prime$ & $SU(2)_L$ & $U(1)_{R}$\\
\hline
\hline
$\zeta_R$ & $\square$ & $\mathbf{4}$ & $\mathbf{1}$ & $\mathbf{1}$ & $-\frac{1}{2N_{\rm HC}}$ \\
$\zeta_L$ & $\square$ & $\mathbf{1}$ & $\mathbf{4}$ & $\mathbf{1}$ & $-\frac{1}{2N_{\rm HC}}$ \\
\hline
$\chi^i_L$ & $\mathbf{1}$ & $\mathbf{4}$ & $\mathbf{1}$ & $\mathbf{2}$ & $0$  \\
$\chi^i_R$ & $\mathbf{1}$ & $\mathbf{1}$ & $\mathbf{4}$ & $\mathbf{2}$ & $0$  \\
\hline
$\psi^3_L$ & $\mathbf{1}$ & $\mathbf{4}$ & $\mathbf{1}$ & $\mathbf{2}$ & $0$ \\
$\psi^{3u}_R$ & $\mathbf{1}$ & $\mathbf{4}$ & $\mathbf{1}$ & $\mathbf{1}$ & $\frac{1}{2}$ \\
$\psi^{3d}_R$ & $\mathbf{1}$ & $\mathbf{1}$ & $\mathbf{4}$ & $\mathbf{1}$ & $-\frac{1}{2}$ \\
$\psi^{1,2}_L$ & $\mathbf{1}$ & $\mathbf{1}$ & $\mathbf{4}$ & $\mathbf{2}$ & $0$ \\
$\psi^{1,2;u}_R$ & $\mathbf{1}$ & $\mathbf{1}$ & $\mathbf{4}$ & $\mathbf{1}$ & $\frac{1}{2}$ \\
$\psi^{1,2;d}_R$ & $\mathbf{1}$ & $\mathbf{1}$ & $\mathbf{4}$ & $\mathbf{1}$ & $-\frac{1}{2}$ \\
\hline
\end{tabular}
\end{center}
\begin{center}
\renewcommand{\arraystretch}{1.2}
\begin{tabular}{|c|c|c|c|c|c|c|}
\hline
Field & $SU(N_{\rm HC}+1)$ & $SU(4)$ & $SU(4)^\prime$ & $SU(2)_L$ & $SU(2)_R$\\
\hline
\hline
$\Xi_R=(\zeta_R,\psi^{3u}_R)$ & $\square$ & $\mathbf{4}$ & $\mathbf{1}$ & $\mathbf{1}$ & $\mathbf{1}$ \\
$\Xi_L=(\zeta_L,U_L)$ & $\square$ & $\mathbf{1}$ & $\mathbf{4}$ & $\mathbf{1}$ & $\mathbf{1}$ \\
\hline
$\chi^i_L$ & $\mathbf{1}$ & $\mathbf{4}$ & $\mathbf{1}$ & $\mathbf{2}$ & $\mathbf{1}$  \\
$\chi^i_R$ & $\mathbf{1}$ & $\mathbf{1}$ & $\mathbf{4}$ & $\mathbf{2}$ & $\mathbf{1}$  \\
\hline
$\psi^3_L$ & $\mathbf{1}$ & $\mathbf{4}$ & $\mathbf{1}$ & $\mathbf{2}$ & $\mathbf{1}$\\
$\psi^{3}_R=(U_R,\psi^{3d}_R)$ & $\mathbf{1}$ & $\mathbf{1}$ & $\mathbf{4}$ & $\mathbf{1}$ & $\mathbf{2}$ \\
$\psi^{1,2}_L$ & $\mathbf{1}$ & $\mathbf{1}$ & $\mathbf{4}$ & $\mathbf{2}$ & $\mathbf{1}$ \\
$\psi^{1,2}_R$ & $\mathbf{1}$ & $\mathbf{1}$ & $\mathbf{4}$ & $\mathbf{1}$ & $\mathbf{2}$ \\
\hline
\end{tabular}
\end{center}
\caption{4321 deconstruction (top) and possible UV completion (bottom). To cancel the anomalies, we need $N_{\rm HC}=2N_{\chi}+1$. At some high scale, $SU(N_{\rm HC}+1)\to SU(N_{\rm HC})\times U(1)_{\rm HC}$, and $U(1)_{\rm HC}\times SU(2)_R\to U(1)_X$. At this scale, $U_L$ and $U_R$ can get a mass and be integrated out, recovering the top table.}
\label{tab:SU4Deconstruction}
\end{table}

A perhaps more natural choice for fermionic charges would be to isolate the top quark by locating $\psi_L^3=(q_L^3\;\ell_L^3)$ and $\psi_R^{3u}=(t_R\;\nu_R^3)$ in the first site and $\psi_R^{3d}=(b_R\;\tau_R)$, together with first- and second-family fermions, in the second. This choice gives rise to a $U(2)_q\times U(2)_u\times U(3)_d\times U(2)_\ell\times U(3)_e$ flavor symmetry in the gauge sector, which coincides with the approximate symmetry of the Yukawa sector and thus offers a good starting point to explain its structure dynamically.\footnote{The minimal set and size of the spurions breaking this symmetry that are required to accommodate the SM Yukawa structure is discussed in~\cite{Greljo:2022cah}.} As before, it is possible to argue that the Higgs field is located in the first site and thus only the top Yukawa is generated in first approximation, whereas all other Yukawas are generated through subleading mass mixing effects. A dedicated study of the corresponding flavor symmetry spurions as well as their possible dynamical origin is beyond the scope of this work and will be discussed elsewhere.
Possible gauge anomalies are more clearly seen 
using the bigger global group 
\begin{equation}
\mathcal{G}_{4421}=SU(4)\times SU(4)^{\prime}\times SU(2)_L\times U(1)_R\supset \mathcal{G}_{4321},
\end{equation}
where $SU(2)_L\times U(1)_R$ acts universally. Anomaly cancellation of the gauge group $\mathcal{G}_{4321}$ will be inherited from anomaly cancellation of $\mathcal{G}_{4421}$.
Cubic and gravitational anomalies of $U(1)_R$ cancel like in the SM because $U(1)_R$ is universal. 
Cubic anomalies of $SU(4)$ or $SU(4)^{\prime}$ are canceled by choosing appropriately the number of will-be-vector-like fermions, $N_{\rm HC}=2N_{\chi}+1$.
The splitting of top and bottom fermions also causes mixed anomalies between $SU(4)^{\prime}$ or $SU(4)$ and $U(1)_R$:
\begin{equation}
\sum_{\psi \in s}  q^{(R)}_{\psi} = 
-\sum_{\psi \in s^{\prime}} q^{(R)}_{\psi}=
\frac{1}{2},\label{eq:4321MixedAnomaly}
\end{equation}
where $s$ ($s^{\prime}$) represents the set of elementary fields charged under $SU(4)$ ($SU(4)^{\prime}$) and $q^{(R)}_{\psi}$ is their charge under $U(1)_R$ (all of them are RH fermions).
However, these anomaly contributions can be compensated by charging hyper-quarks under $U(1)_R$, or equivalently, $U(1)_X$, following a similar structure to that of toy model~II. 
The complete implementation is described in \cref{tab:SU4Deconstruction} (top). 
Thus, hyper-quarks also contribute to the sums of \cref{eq:4321MixedAnomaly}, making them vanish. Furthermore, hyper-quarks do not create other $U(1)_R$ anomalies due to their vector-like character under $U(1)_R$.

As with toy model II, it is possible to extend this model to a larger UV symmetry where the $U(1)$ symmetry is embedded into other $SU(N)$ factors, see \cref{tab:SU4Deconstruction} (bottom).
In this case, the right-handed top appears as the hyper-lepton associated to the hyper-quarks in the hyper-Pati-Salam extension. If the Higgs is realized as a pseudo-Nambu-Goldstone boson by extending the composite sector in a similar manner to~\cite{Fuentes-Martin:2020bnh}, four-fermion operators that induce the top Yukawa coupling after confinement could be generated by the same NP responsible of breaking $SU(N_{\rm HC}+1)\times SU(2)_R\to SU(N_{\rm HC})\times U(1)_X$. Depending on the details of this NP sector, which should lie around the $10$~TeV scale, the largeness of the top-Yukawa compared to the others Yukawas could be dynamically explained given the different charges of the right-handed top in this UV completion.
Notice that this completion with a semisimple group makes the cancellation of mixed anomalies of our 4321 model more transparent, in a similar way than the PS completion does for the SM.

At energies below the confinement scale, the hyper-sector is better described by an effective chiral action of the would-be Goldstone bosons containing the WZW terms of~\cref{sec:EffChi}. Identifying $SU(4)$ with $SU(N)_R$ and $SU(4)^{\prime}$ with $SU(N)_L$, the effective description for all these 4321 models has the would-be-Goldstone bosons of the coset $SU(4)_L\times SU(4)_R/SU(4)_V$, all of them eaten by the gauge bosons. With them we can write the pure WZW term of~\cref{eq:WZWSUN3} and, for the model of \cref{tab:SU4Deconstruction} (top), also the mixed WZW term of \cref{eq:WZWSUN2U1,eq:WZWSUN2U14d} with $\bar m= -1/2$ and $U(1)\equiv U(1)_R$. 
Since only $\mathcal{G}_{4321}$ is gauged, the gauge fields appearing on these WZW terms should be taken to be
\begin{equation}
A_R = \sum_{a=1}^{15} A_{4}^{a}\,T_a,~~~~
A_L = \sum_{a=1}^{8} A_{3}^{a}\,T_a + 
\sqrt{\frac{2}{3}}\, A_X\,T_{15} ,
\label{eq:gaugingLR}
\end{equation}
and, for the model of \cref{tab:SU4Deconstruction} (top), also
\begin{equation}
A_{U(1)} =\,A_X,\label{eq:gaugingU1}
\end{equation}
where $A_4^{1,\ldots,15}$, $A_3^{1,\ldots,8}$ and $A_X$ correspond to the gauge fields of $SU(4)$, $SU(3)$ and $U(1)_X$ respectively, and $T_a$ are the usual $SU(4)$ generators with the canonical normalization ${\rm Tr} (T_a T_b) = \frac{1}{2}\, \delta_{ab}$, so $T_{a}$ with $a=1,\ldots,8$ expand $SU(3)$ and $T_{15}=\frac{1}{2\sqrt{6}}$diag$(1,1,1,-3)$.

Holographic realizations are also possible if we include the CS terms as in \cref{sec:Hologr}.
Elementary fermions are then localized in the UV brane, and the 5d bulk with the geometry of \cref{eq:AdS5} effectively describes the composite sector.
Its global symmetries, $SU(4)_L\times SU(4)_R\times U(1)_V$ where $SU(4)_R\equiv SU(4)$ and $SU(4)_L\equiv SU(4)^{\prime}$, are implemented in the holographic description as gauge symmetries in the 5d bulk.
As for the boundary conditions of these 5d gauge fields, while in the IR brane they describe the appearance of a condensate breaking the symmetry $SU(4)_L\times SU(4)_R\to SU(4)_V$, $A_L = A_R$, in the UV brane they are dictated by the gauging and are given by \cref{eq:gaugingLR} and also \cref{eq:gaugingU1} with $U(1)\equiv U(1)_V$ for the model of \cref{tab:SU4Deconstruction}~(top).\footnote{These boundary conditions apply only to the 4d components of the gauge fields.}
The 5d actions for all the 4321 models discussed have the CS forms of \cref{eq:HoloCS,eq:HoloMixCS}.
They cancel all anomaly contributions of the elementary fermions in the UV brane. In particular, for the model of \cref{tab:SU4Deconstruction}~(top), the CS form of \cref{eq:HoloMixCS} cancels the mixed anomalies. Using the identification $U(1)\equiv  U(1)_R$, it requires $\bar m=-1/2$.

Regarding the phenomenological implications of these models, all 4321 implementations discussed here predict a massive vector leptoquark that is able to address the tensions observed in $B$-decays~\cite{Capdevila:2023yhq,Koppenburg:2023ndc}, particularly in the $R_{D^{(*)}}$ measurements~\cite{Aebischer:2022oqe}. However, an added advantage of the one where $b_R$ and $\tau_R$ are charged under $SU(4)^{\prime}$ (c.f. \cref{tab:SU4Deconstruction}) is that the absence of leptoquark couplings to the right-handed bottom quark weakens the bounds from LHC searches~\cite{Baker:2019sli,Haisch:2020xjd}. This implementation further predicts $R_{D}/R_{D}^{\rm SM}= R_{D^{*}}/R_{D^{*}}^{\rm SM}$, which might be experimentally testable if the anomaly persists.

\subsection{\texorpdfstring{$SU(2)_L$}{SU(2)L} deconstructions}

A similar approach can be used for deconstructing other simple factors of the SM gauge group. For instance, the deconstruction of $SU(2)_L$ has been extensively discussed (see for instance~\cite{Li:1981nk,Muller:1996dj,Malkawi:1996fs,Chiang:2009kb,Hsieh:2010zr,Fuentes-Martin:2014fxa,Boucenna:2016wpr,Boucenna:2016qad,Davighi:2023xqn,Capdevila:2024gki}). This is, the extension of $G_{\rm SM}$ to $SU(3)_c\times SU(2)_l\times SU(2)_h\times U(1)_Y$, where $SU(2)_l$ charges light families and $SU(2)_h$ the third family.\footnote{UV completions of this deconstruction such that the different $SU(2)$ factors unify in a simple group have been studied in~\cite{Davighi:2022fer}.} This deconstruction realizes the $U(2)_q\times U(2)_u\times U(3)_d\times U(2)_\ell\times U(3)_e$ flavor symmetry, which only allows for third-family Yukawas at the renormalizable level if the Higgs is charged under $SU(2)_h$. However, as pointed out in~\cite{Antusch:2023shi}, this flavor symmetry suggests a wrong pattern for the PMNS matrix because it imposes selection rules on the Weinberg operator. 
We can then be less ambitious and use $SU(2)_L$ deconstructions to address the flavor hierarchies only in the quark sector.
Keeping the same structure for quarks but charging all lepton doublets under the same group factor would promote $U(2)_{\ell}$ to $U(3)_{\ell}$ as in the SM.\footnote{Other possibilities to fix this issue could be adding heavy neutral leptons to implement an inverse see-saw mechanism like in~\cite{Fuentes-Martin:2020pww}.} Such models do not address the hierarchy between the $\tau$ and light-lepton masses, but can have interesting phenomenological implications~\cite{Capdevila:2024gki}.
Of course, like in toy model~II, these charge assignments create mixed anomalies between $SU(2)_{l,h}$ and $U(1)_Y$, or more specifically, the $U(1)_{B-L}$ component of hypercharge. Then, we can use them to illustrate how the anomaly-cancellation mechanism is implemented.
For concreteness, we will assume that we want to charge all leptons under $SU(2)_h$ (other possibilities can be built following the same logic).\footnote{For instance, another interesting possibility is charging all leptons under $SU(2)_l$, that would explain an overall suppression of the lepton Yukawas.} 
Let us say the breaking $SU(2)_{l}\times SU(2)_{h}\to SU(2)_L$ is triggered by a condensate of a strong sector with two flavors of hyper-quarks in the fundamental representation of $SU(N_{\rm HC})$. If right-handed hyper-quarks are arranged into doublets of the $SU(2)_l$ symmetry, and left-handed hyper-quarks into doublets of $SU(2)_h$, all local anomalies are canceled provided the hyper-quarks also carry hypercharge $1/N_{\rm HC}$ (see \cref{tab:fieldcontentSU(2)} (top)). Moreover, global anomalies require $N_{\rm HC}$ to be even. Interestingly, this condition ensures that the hyper-baryons of the strongly-coupled sector have integer electric charge: they will fit in real representations of $SU(2)_L$ (and have hypercharge 1).

As it happened with the 4321 deconstruction, we can UV-complete the model to explain the fractional charges of the hyper-quarks. Let us take $SU(N_{\rm HC}+2)\times SU(3)_c\times SU(2)_l\times SU(2)_h\times U(1)^{\prime}_Y$ and arrange the fields as in~\cref{tab:fieldcontentSU(2)} (bottom). Then, at some high scale above the confinement scale of $SU(N_{\rm HC})$, $SU(N_{\rm HC}+2) \times U(1)^{\prime}_Y$ breaks to $SU(N_{\rm HC})\times U(1)_Y$, with $U(1)_Y=U(1)_{\rm HC} \oplus U(1)_Y^{\prime}$ and the charges of $\{\zeta_{L,R}\}$ and $\{\ell_L^{1,2},L_R^{1,2}\}$ under $U(1)_{\rm HC}\subset SU(N_{\rm HC}+2)$ being $1/N_{\rm HC}$ and $-1/2$, respectively. Furthermore, $L_L^{1,2}$ and $L_R^{1,2}$ form a Dirac pair after this breaking and can naturally get a mass at the high scale. After integrating out these extra fermions, we recover the model of \cref{tab:fieldcontentSU(2)} (top).

\begin{table}[t]
\begin{center}
\renewcommand{\arraystretch}{1.2}
\begin{tabular}{|c|c|c|c|c|c|c|}
\hline
Field & $SU(N_{\rm HC})$ & $SU(2)_h$ & $SU(2)_l$ & $U(1)_Y$\\
\hline
\hline
$\zeta_R$ & $\square$ & $\mathbf{1}$ & $\mathbf{2}$ & $\frac{1}{N_{\rm HC}}$ \\
$\zeta_L$ & $\square$ & $\mathbf{2}$ & $\mathbf{1}$ & $\frac{1}{N_{\rm HC}}$  \\
\hline
$\ell^{1,2,3}_L$ & $\mathbf{1}$ & $\mathbf{2}$ & $\mathbf{1}$ & $-\frac{1}{2}$ \\
$q^{3}_L$ & $\mathbf{1}$ & $\mathbf{2}$ & $\mathbf{1}$ & $\frac{1}{6}$ \\
$q^{1,2}_L$ & $\mathbf{1}$ & $\mathbf{1}$ & $\mathbf{2}$ & $\frac{1}{6}$ \\
\hline
\end{tabular}
\end{center}
\begin{center}
\renewcommand{\arraystretch}{1.2}
\begin{tabular}{|c|c|c|c|c|c|c|}
\hline
Field & $SU(N_{\rm HC}+2)$ & $SU(2)_h$ & $SU(2)_l$ & $U(1)_Y^{\prime}$\\
\hline
\hline
$\Xi_R=(\zeta_R,L^1_R,L^2_R)$ & $\square$ & $\mathbf{1}$ & $\mathbf{2}$ & $0$ \\
$\Xi_L=(\zeta_L,\ell^1_L,\ell^2_L)$ & $\square$ & $\mathbf{2}$ & $\mathbf{1}$ & $0$  \\
\hline
$L^{1,2}_L$ & $\mathbf{1}$ & $\mathbf{1}$ & $\mathbf{2}$ & $-\frac{1}{2}$ \\
\hline
$\ell^{3}_L$ & $\mathbf{1}$ & $\mathbf{2}$ & $\mathbf{1}$ & $-\frac{1}{2}$ \\
$q^{3}_L$ & $\mathbf{1}$ & $\mathbf{2}$ & $\mathbf{1}$ & $\frac{1}{6}$ \\
$q^{1,2}_L$ & $\mathbf{1}$ & $\mathbf{1}$ & $\mathbf{2}$ & $\frac{1}{6}$ \\
\hline
\end{tabular}
\end{center}
\caption{Anomalous $SU(2)_L$ deconstruction (top) and UV completion (bottom). Only fields charged under $SU(2)_{l,h}$ are shown. Other charge assignments are the same as in the SM.}
\label{tab:fieldcontentSU(2)}
\end{table}

\subsection{Multiple-factor deconstructions}

So far, we have only considered UV completions where just one simple factor of the SM gauge group is deconstructed, but several group factors can be deconstructed simultaneously. For instance, in \cite{Davighi:2023iks} it has been suggested that the deconstruction of $SU(4)_{PS}\times SU(2)_R$ while $SU(2)_L$ remains universal has particularly interesting properties to explain the SM flavor structure. In those cases, there are several choices for the strongly-coupled sector triggering the breaking:
\begin{itemize}
\item One single representation of hyper-quarks that realizes a hyper-colored flavor symmetry containing all the factors that we want to deconstruct. For instance, to deconstruct $SU(4)_{PS}\times SU(2)_R$, we can have 8 flavors of hyper-quarks realizing $SU(8)_L\times SU(8)_R$ so $SU(4)_{PS}\times SU(2)_R \subset SU(8)_V$.
\item Several complex representations of hyper-quarks, with each realizing as flavor symmetry one of the group factors that we want to deconstruct. In this case, each representation will develop a condensate responsible of the breaking to the diagonal of each factor. 
\item Several strong sectors, each of which responsible for triggering the breaking of one of the deconstructed group factors.
\end{itemize}
In analogy to the model examples described above, it is possible to employ different variants of toy models I and II to implement various anomalous arrangements of the SM fermions in some multiple-factor deconstructions. A dedicated study of these possibilities is however beyond the scope of this paper.

\section{Conclusions}
\label{sec:Conclusions}

In this article, we have proposed new possibilities to build NP models in the context of flavor deconstructions: UV completions of the SM where the gauge group is extended to multiple copies that act non-universally on the fermions. These constructions offer an interesting approach to address the flavour puzzle, as they forbid at the renormalizable level some of the SM Yukawa couplings. These couplings are then generated dynamically from NP contributions at higher scales, thus providing a multi-scale explanation of the flavor hierarchies. Since, at the lowest energy scale, these explanations share most of the accidental global symmetries of the SM Yukawa sector, the corresponding NP at that scale is typically protected from sensitive flavor observables and can lie around the TeV scale.

A new sector that breaks the extended gauge symmetry to the SM one is an essential ingredient in any flavor deconstruction. If this sector consists on scalar fields that acquire a vev and no new fermions are added beside the SM ones, gauge-anomaly cancellation requires charging complete SM families under the same factors. Thus, when the deconstructed group is (semi)simple, the splitting of SM families among the deconstructed factors typically creates mixed anomalies between these factors and some $U(1)$ group related to hypercharge. In this article, we have shown that, if the breaking is triggered by the condensate of a new strongly coupled hyper-sector, the fundamental fermionic degrees of freedom of the hyper-sector (hyper-quarks) can be charged fractionally under this $U(1)$ group in a way that the anomaly-cancellation condition is relaxed. Fermions of the same generation can thus be split among different group factors. We also identified UV completions of these models that avoid fractional charges. These are analogous to Pati-Salam unification, but now applied to the new hyper-sector. Besides relaxing the anomaly-cancellation condition, other advantages of using a strongly-coupled sector for the symmetry breaking are their radiative stability and the possibility to extend them to incorporate a composite Higgs, linking the multi-scale picture to solutions of the Higgs hierarchy problem. We have provided the fundamental description of these models in terms of hyper-quarks, but also reviewed the cancellation of anomalies in their effective chiral description through WZW terms and in holographic realizations. Interestingly, anomaly cancellation in the chiral description and in the holographic picture share some formal resemblance, with Chern-Simons forms playing similar roles. 

The application of these ideas open new ways to build novel and well-motivated SM deconstructions. As a practical example, we concentrated on the explanation of the hierarchy between top and bottom quarks, which normally remains unexplained by standard deconstructions. With the new mechanisms explored in this paper, models with a gauge structure that uniquely identify the top quark from other SM fermions become possible and natural. Following this logic, a 4321 model featuring quark-lepton unification of the third family, but excluding the right-handed bottom and tau fields has been proposed. Other possibilities that we have discussed are an $SU(2)_L$ deconstruction breaking universality in the quark but not in the lepton sector, or the deconstruction of multiple semisimple groups.

To conclude, deconstructing flavor anomalously offers the possibility of exploring NP models at the TeV scale that realise different flavor symmetries than those found in standard flavor deconstructions. In some cases, these could be more convenient to describe the SM flavor patterns. For instance, a gauge sector realizing a $U(2)_q\times U(2)_u\times U(3)_d$ flavor symmetry in the quark sector naturally addresses the top-bottom hierarchy and could be a better starting point for a multi-scale explanation of the flavor hierarchies.

\acknowledgments

We would like to thank Joe Davighi and Nud{\v z}eim Selimovi{\'c} for useful discussions. We further thank Joe Davighi for helpful comments on the manuscript. The work of JFM is supported by the Spanish Ministry of Science and Innovation (MCIN) and the European Union NextGenerationEU/PRTR under grant IJC2020-043549-I, by the MCIN and State Research Agency (SRA) projects PID2019-106087GB-C22 and PID2022-139466NB-C21 (ERDF), and by the Junta de Andaluc\'ia projects P21\_00199 and FQM101. The work of JML has been supported by the European Research Council (ERC) under the European Union’s Horizon 2020 research and innovation program under grant agreement 833280 (FLAY) in the initial stages of the project, and by the grant CSIC-20223AT023 in the final ones. JML also acknowledges the support of the Spanish Agencia Estatal de Investigacion through the grant “IFT Centro de Excelencia Severo Ochoa CEX2020-001007-S”.


\bibliographystyle{JHEP}
\bibliography{references}

\end{document}